%% file: mdetlsst.tex
\documentclass[twocolumn,twocolappendix,astrosym]{openjournal}

\usepackage{graphicx}

\usepackage{latexsym,amssymb}

\usepackage{amsmath,morefloats}
\usepackage[backref,breaklinks,colorlinks,citecolor=blue]{hyperref}
\usepackage{natbib,graphicx,amsmath,subfigure,color}
\usepackage[usenames,dvipsnames]{xcolor}
\usepackage{verbatim}
\usepackage{threeparttable}

\usepackage{pgfplots}
\pgfplotsset{compat=1.17}
\usepgfplotslibrary{fillbetween}
\usepgfplotslibrary{groupplots}
\usetikzlibrary{patterns}

\definecolor{bisque}{HTML}{ffe8c4}
\definecolor{tan}{HTML}{d2b48c}
\definecolor{sand}{HTML}{e1bf92}
\definecolor{sand1}{HTML}{f6d7b0}
\definecolor{sand2}{HTML}{f2d2a9}
\definecolor{sand3}{HTML}{eccca2}
\definecolor{sand4}{HTML}{e7c496}
\definecolor{sand5}{HTML}{e1bf92}

\definecolor{desert1}{HTML}{d9a76c}
\definecolor{desert2}{HTML}{a48963}
\definecolor{desert3}{HTML}{674c37}
\definecolor{desert4}{HTML}{512f26}
\definecolor{desert5}{HTML}{211813}

\definecolor{bsand1}{HTML}{b18e5d}
\definecolor{bsand2}{HTML}{97744f}
\definecolor{bsand3}{HTML}{590d0d}
\definecolor{bsand4}{HTML}{3b0808}
\definecolor{bsand5}{HTML}{1d0808}

\newcommand{\descwl}{\texttt{WeakLensingDeblending}}

\newcommand{\vecg}{\mbox{\boldmath $g$}}

\newcommand{\vecc}{\mbox{\boldmath $c$}}
\newcommand{\vecm}{\mbox{\boldmath $m$}}

\newcommand{\galsim}{\texttt{GALSIM}}
\newcommand{\ngmix}{\texttt{ngmix}}

\newcommand{\snr}{$S/N$}
\newcommand{\Tratio}{$T/T_{PSF}$}

\newcommand{\calexp}{\texttt{Exposure}}

\newcommand{\dm}{LSST Science Pipelines}

\newcommand{\mcal}{\textsc{metacalibration}}
\newcommand{\mdet}{\textsc{metadetection}}

\newcommand{\Mdet}{\textsc{Metadetection}}

\newcommand{\vecgam}{\mbox{\boldmath $\gamma$}}
\newcommand{\mthresh}{0.002}
\newcommand{\mshear}{0.02}
\newcommand{\fullarea}{41,600}

\newcommand{\vonkarman}{{von K\'arm\'an}~}

\shorttitle{Metadetection for Rubin}
\shortauthors{Sheldon, Becker, Jarvis, Armstrong}

\begin{document}


\title{Metadetection Weak Lensing for the Vera C. Rubin Observatory}

\author{Erin S. Sheldon}
\affil{Brookhaven National Laboratory, Bldg 510, Upton, New York 11973, USA}
\author{Matthew R. Becker}
\affil{High Energy Physics Division, Argonne National Laboratory, Lemont, IL 60439, USA}
\author{Michael Jarvis}
\affil{Department of Physics and Astronomy, University of Pennsylvania, Philadelphia, PA 19104, USA}
\author{Robert Armstrong}
\affil{Lawrence Livermore National Laboratory, Livermore, CA 94551, USA}
\author{The LSST Dark Energy Science Collaboration}

\begin{abstract}

        Forthcoming astronomical imaging surveys will use weak gravitational
        lensing shear as a primary probe to study dark energy, with accuracy
        requirements at the 0.1\% level.  We present an implementation of the
        \mdet\ shear measurement algorithm for use with the Vera C. Rubin
        Observatory Legacy Survey of Space and Time (LSST).  This new code
        works with the data products produced by the \dm, and uses the pipeline
        algorithms when possible.  We tested the code using a new set of
        simulations designed to mimic LSST imaging data.  The simulated images
        contained semi-realistic galaxies, stars with representative
        distributions of magnitudes and galactic spatial density, cosmic rays,
        bad CCD columns and spatially variable point spread functions.  Bright
        stars were saturated and simulated ``bleed trails'' were drawn.
        Problem areas were interpolated, and the images were coadded into small
        cells, excluding images not fully covering the cell to guarantee a
        continuous point spread function.  In all our tests the measured shear
        was accurate within the LSST requirements.



\end{abstract}

\section{Introduction} \label{sec:intro}

New astronomical imaging surveys coming online in the next few years will use
weak gravitational lensing shear as a primary probe to study dark energy.
These ``stage IV'' dark energy experiments, the Rubin Observatory Legacy Survey
of Space and Time \citep[LSST,][]{IvezicLSST2008}, the Euclid mission
\citep{Euclid2011} and surveys with the Nancy Grace Roman Space Telescope
\citep{Roman2015,AkesonRoman2019} require shear measurement methods that are
accurate to a few tenths of a percent \citep{Massey2013,SRD}, and
implementations of those algorithms that work with the data products of each
survey.

A number of promising shear measurement techniques have been developed over the
last few years.  The \mdet\ shear measurement method \citep{mdet20} has better
than 0.1\% accuracy for noisy data, overlapping, or ``blended'', galaxy images
and a constant applied shear
\citep{mdet20,HoekstraMdet2021a,HoekstraMdet2021b}.  At the time of writing,
other recently developed methods provide similar accuracy for isolated galaxy
images \citep{BernBFD2016, LiFPFSBlending2022} but require further percent
level corrections from simulations to account for blending effects
\citep{mdet20,LiFPFSBlending2022}\footnote{Such simulations may be required in
any case to jointly calibrate the redshift dependent shear and redshift
distribution of the source galaxies \citep{MacCrann2022,LiNofz2022}, but
smaller biases can be corrected more reliably.}.

Real astronomical images contain additional features that must be addressed in
order to obtain accurate shear measurements.  A typical image may be expected
to contain, among other features, cosmic rays, bad CCD columns, saturation,
Milky Way stars (which are not lensed significantly) and star ``bleed trails''.
For computational efficiency the images may be remapped and summed into
aggregates, or ``coadded'', which can result in spatially correlated image
noise.

An accurate shear measurement method must be unbiased (up to higher order shear
effects) in the presence of these image features.  That is, if the input data
are well characterized in terms of flux calibration, astrometry, background
determination, noise properties, and PSF, and if problematic features are
identified and masked, the technique should provide an accurate shear
measurement.

In actual analyses on real data, errors in the calibration and characterization
of the data may contribute significantly to the systematic error in the shear
measurement.  For an accurate shear measurement, the ultimate limit may be the
characterization of the data, not the method itself.

In this work we demonstrate that the \mdet\ technique can accurately calibrate
shear estimates in such featureful, but well-characterized data.  We simulate
images containing galaxies, stars with a realistic galactic density
distribution, spatially variable PSF, and proxies for the above image artifacts
(see \S \ref{sec:sim}), which we interpolate, or ``warp'', onto a common
reference frame and sum into overlapping coadds (see \S \ref{sec:coadding}).
Each feature was switched on or off independently to control potential sources
of error.  We simulate the expected data products produced by the \dm\
\citep{BoschLSST2019,BoschHSC2017}\footnote{\url{https://github.com/lsst}}, and
use a new \mdet\ code designed specifically to work with Science Pipeline data
structures and algorithms\footnote{The software used in this work is freely
available on the internet.  URLs are provided as footnotes to the text.}.  We
describe \mdet\ in \S \ref{sec:mdet}, and the results in \S \ref{sec:results}.

\section{Simulation Features} \label{sec:sim}

The basic simulation was similar to that created for \citep{mdet20}.  We added
additional features such as image artifacts, cosmic rays, stars, and saturated
stars which we will describe in the following sections.  All images were
rendered using the \galsim\ python
package\footnote{\url{https://github.com/GalSim-developers/GalSim}}. For this work,
we used the newly written simulation package
\texttt{descwl-shear-sims} version
0.4.2\footnote{\url{https://github.com/LSSTDESC/descwl-shear-sims}}.

As our basic data product, we simulated the calibrated exposure images produced
by the \dm\ version \texttt{w\_2021\_32}.  This data is stored in an \calexp\
data structure, which carries a calibrated, background subtracted image, along
with an estimated noise variance image plane, world coordinate system
transformation (WCS) and position-dependent PSF model.  Problem areas
associated with saturation, cosmic rays, and bad columns are marked as separate
bits in an integer bit mask image plane.


Other than simple, constant background estimation errors, which were easy to
implement and test (see \S \ref{sec:sim:bgerr}), we did not simulate scenarios
in which the input data were miscalibrated.  For example we did not test the
consequences of inaccurate PSF models, noise estimates or photometric
calibrations.  Our motivation was to test the performance of \mdet\ using
well-characterized data. It may be necessary to propagate or simulate the
effects of miscalibrated data in order to characterize the final shear
calibration in real data analyses.

Importantly, our approach to simulations in this work was not to simulate every
possible effect in detail. Such simulations, like the Data Challenge 2 (DC2)
simulations \citep{DC2Abolfathi2021} for example, while physically more
realistic than our approach here, are exceedingly complex to analyze and can be
computationally much more expensive than the approach we took in this work.
Instead, we approached our simulation analysis as a series of detailed
numerical experiments where single features are changed one at a time. This
level of control combined with reasonably fast codes was needed for us to reach
our goal of constraining the bias in our algorithms with a precision better
than the LSST requirements of \mthresh\ (see \S \ref{sec:lsstreq}). For
example, for the tests shown in \S \ref{sec:results:full}, we simulated a total
of \fullarea\ square degrees, providing constraints well within our
requirements at 99.7\% confidence. On the other hand, the simulated area in DC2
is about 300 square degrees, which would provide constraints only at the
percent level. Furthermore, using the existing DC2 simulation does not allow
control over the simulation features, so that in general it would not be
possible to isolate the source of shear measurement biases. Future work
building on our approach here and carefully examining one feature at a time
will be essential for delivering accurate shear measurements for LSST.

\subsection{Image Filters and Noise} \label{sec:sim:noise}

We used the \descwl\ package
\citep{DESCWLSanchez2021}\footnote{\url{https://github.com/LSSTDESC/WeakLensingDeblending}}
to predict the image noise according to the Rubin Observatory/LSST filter being
simulated.  In all simulation runs we used noise $n$ for the final 10 year
coadd.  If $N$ simulated images were created, representing data from individual
exposures to be coadded, we re-scaled the noise appropriately, such that the
noise in each image was $n * \sqrt{N}$.

We did not simulate a realistic background and background noise for each image
(see \S \ref{sec:sim:bgerr}).  We instead used Gaussian noise to approximate
the Poisson noise that would remain in each image after background subtraction.
After drawing all image features and adding noise, we re-scaled the images to a
common zero point of 30 before coadding.

We did not include the Poisson noise from objects.  This was mainly an
optimization, as it is inefficient to simulate the large number of photons from
bright objects.   We have tested \mcal\ with object Poisson noise for faint,
sky noise dominated objects, but not brighter objects, which we leave to future
work.

\subsection{Image World Coordinate System, Rotations and Dithers} \label{sec:sim:rotdith}

Each image was simulated as a tangent plane projection of the sky onto the
image frame with the expected LSST camera pixel scale of 0.2 arcseconds
\citep{IvezicLSST2008}.  The world coordinate system (WCS) transformation between
sky and pixel coordinates was represented as a \galsim\ tangent plane
WCS object (\texttt{TanWCS}).
Each simulated WCS transformation had a random rotation applied, in order to
mock up the camera rotations used by LSSTCam \citep{IvezicLSST2008}.  The
center of each image was shifted, or ``dithered'', randomly in two dimensions
relative to the center of the desired coadded image.  For efficiency, we used
dithers that were unrealistically small, within two pixels, to minimize portions of the
image that would not overlap the final coadd.  We created images with sizes just
large enough that the rotated image, after dithers, would have no edge
crossing the coadd region, again to avoid waste:  coadding an image with an
edge produces a discontinuous PSF in the final coadd, so such images would be
discarded.

The simulated image was copied into an \calexp\ data structure with an
equivalent \dm\ tangent plane \texttt{SkyWCS}.

\subsection{Point Spread Function} \label{sec:sim:psfs}

For our basic PSF model we used a Moffat profile \citep{Moffat1969} with shape
parameter $\beta=2.5$ and full width at half maximum (FWHM) of 0.8 arcseconds.

We also generated spatially variable PSF models, varying both the ellipticity
and size across the image, using the methods described in Appendix~A of
\citet{mdet20} based on work by \citet{heymans2012}.  \citet{heymans2012} used
images with high stellar density to fit a \vonkarman\ model of atmospheric
turbulence to the PSF variation. \citet{mdet20} used this model, with a
modification to reduce unrealistically large power below 1'', to generate
realizations of spatially variable PSFs using Gaussian random fields. The PSF
model was again a Moffat profile with shape parameter $\beta=2.5$ but with
variable size and shape.

In \citet{mdet20} this approximate variable PSF model was compared to detailed
atmospheric and optics simulations for the Dark Energy Survey (DES), and the
variation was tuned to exceed the expectations of real data by a factor of 10.
We used the same models for this work, but with a variation tuned to match,
rather than exceed, expectations for DES.  The level of variations for DES at
the Cerro Tololo site should be a rough approximation to the variations at the
nearby Rubin observatory at Cerro Pachón.  The median FWHM of the generated
PSFs was 0.8 arcseconds for all bands.  We also ran simulations with larger
than expected variations in order to test the accuracy of the PSF coadd under
extreme, unrealistic conditions.

We did not include other sources of spatial PSF variation such as effects from
the optical surfaces in the telescope.  We also did not include any
chromaticity in the PSF \citep{PlazasPSF2012,MeyersBurchat2015,Kamath2020} or
galaxy color gradients that complicate the PSF correction.  We leave such
considerations to future work.

\subsection{Stars} \label{sec:sim:stars}

We simulated stars using fluxes and Milky Way densities sampled from the LSST
DESC Data Challenge Two (DC2) simulation catalogs \citep{DC2Abolfathi2021}.
For each simulated field we sampled randomly, with replacement, from the map of
stellar density used to generate DC2, rejecting densities higher than 100 per
square arcminute.  This density represents the total number of stars drawn, not
the number detected.  The multi-band flux for each star was sampled with
replacement from the DC2 star catalog.  We modeled each star as a point
source convolved with the point spread function
(see \S \ref{sec:sim:psfs}).  Stars were allowed to saturate and have an
associated bleed trail (see \S \ref{sec:sim:satbleeds}).

\subsection{Image Saturation, Star Masking and Star Bleed Trails} \label{sec:sim:satbleeds}

The value in each pixel was artificially limited in order to simulate
saturation, saturated pixels were marked with an appropriate flag in the
integer bitmask image of the \calexp, and finally the variance for those
pixels was set to infinity.  Non-linear detector response was not simulated.

Saturated stars were over-drawn with a simulated bleed trail image, taken from
a set of pre-generated templates identified in images of bright stars created
using the DC2 code.  For each saturated star in our simulation, we found a star
in the template set with closely matched flux in the filter of interest and
drew the associated bleed image directly over the star image with a value set
to the saturation level.  The bleed pixels were marked with the appropriate
flag in the bitmask.

Before performing PSF deconvolution on the coadds, we further masked
saturated stars with a circle that covered the star out to the radius
where the profile reached the noise level.  Note in real data such a mask would
need to be determined algorithmically. This region was set to zero in the image
and marked appropriately in the bitmask.   This mask does not necessarily cover
the bleed trail completely, although the trails were interpolated in the
original images before warping (see \S \ref{sec:coadding}).  We find that these
unmasked trails do not cause a shear bias, which we attribute to the camera
rotations that randomize the direction of the trail on the sky.

Masks with sharp features can cause ringing in the FFTs used by the \mdet\
algorithm.   Each star mask is like a circular ``tophat'', which will have a
sharp feature where it intersects an object in the image.  We mitigated this
effect using the apodization procedure described in \citet{BeckerMdetCoadd}.
We extended the star mask by 16 pixels and forced the mask to smoothly
transition from zero in the interior of the circle to unity at the expanded
edge. The transition region was parameterized using the cumulative integral of
a triweight kernel. This kernel is a function of two parameters, $m$ and $h$,
and is defined for a point $x$ with quantity $y = (x-m)/h$ as
\begin{equation}
K(x, m, h) = \begin{cases}
0 & y < -3 \\
(-5y^7 / 69984 \\
+ 7y^5 / 2592 \\
- 35y^3 / 864 & -3 \le y \le 3 \\
+ 35y / 96 \\
+ 1 / 2) \\
1 & y > 3
\end{cases}
\end{equation}
The kernel goes from zero to unity over a span of $6h$, centered on $m$.
The general logic is to set $h$ large enough that the variation across the edge is
slower than the variation given by the PSF profile, but not so large that area
in the image is wasted unnecessarily. We set $h$ to 1.5 pixels and set $m+3h$ to
the radius of the star mask hole so that the mask reaches unity at its nominal size.

\subsection{Cosmic Rays \& Bad Columns} \label{sec:sim:cosmics_badcols}

We followed \citet{BeckerMdetCoadd} in generating bad column and
cosmic ray artifacts.  For cosmic rays, we selected a random location on the
image, a random angle, and a random length between 10 and 30 pixels. We then
flagged pixels along this line as having been hit by a cosmic ray, making sure
that pixels that touch only along corners have the pixels directly adjacent to
them flagged as well. We set a cosmic ray bit in the bitmask for interpolation
later and set the image value to \texttt{NaN} to ensure that no flagged pixels
are inadvertently used in the final shear estimates.

We generated bad column masks using a slightly modified Monte Carlo generator
from \citet{BeckerMdetCoadd}. Each bad column was a single pixel wide,
positioned randomly on the image. We also added gaps at random to the bad
columns to simulate bad columns which do not span the full CCD.  We generated
a single bad column for each image.

The LSST will generate hundreds of exposures overlapping each location in the
survey.  It is likely that detection and lensing measurements will be carried
out on coadded images, with artifacts interpolated on the original data before
being warped and added.  The aggregate number of artifacts will be large, but
the impact of each artifact will be relatively small.  It was not
computationally feasible to simulate such a large number of individual images
to be coadded in order to test the effect of artifacts in a realistic way.
Rather we simulated a smaller number of images per band, typically 1 but up to
10 (see \S \ref{sec:results:psfvar}), with the noise scaled so that the final
coadd noise matched LSST 10 year data.  Thus, relative to expected LSST data
the impact of each artifact on the coadd was large but the aggregate number of
artifacts was small.

\subsection{Galaxies} \label{sec:sim:galaxies}

We used the \descwl\ package to generate galaxy models
\citep{DESCWLSanchez2021}.  These elliptical, color galaxy models have bulge,
disk and AGN components.  The components have the same morphology in each band.
This catalog has a raw density of $\sim240$ galaxies per square arcminute, and
an effective $i$-band ab magnitude limit of $\sim27$.  Rather than use the
\descwl\ package to draw the models in the image, we drew the models using
Galsim in our code in order to use our own PSF models (see \S
\ref{sec:sim:psfs}), and to allow for more efficient drawing.

We also performed simulation runs with single component, round exponential
galaxies with fixed flux and size.  These models are useful for shear recovery
tests that do not require a realistic galaxy population, but benefit from using
low shape noise shear tracers.  We also used these models for relatively fast
shear tests after major code changes, as further verification after the faster
unit tests had passed.

\subsection{Layouts} \label{sec:sim:layouts}

We ran tests with both random and gridded galaxy layouts.  The random layout
was used for the realistic galaxy population, with a density of objects that
reflected the expected density of galaxies as defined in the \descwl\ package.
The grid was used with the simple, round exponential galaxies to facilitate
relatively fast shear recovery tests.  The grid was square, with spacing
designed to avoid overlap between adjacent galaxies, and thus avoid blending
effects.

See \S \ref{sec:sim:galaxies} for descriptions of each galaxy model type.

\subsection{Image Background} \label{sec:sim:bgerr}

We added a constant to the images to approximate errors in the background
determination.  The motivation was to test if background estimation errors in
individual images can be corrected in the final coadd (see \S
\ref{sec:mdet:detect} and \S \ref{sec:results:more}).

We ran tests with both positive and negative backgrounds, and a wide range of
absolute values, and got consistent results.  For the results presented below,
we chose a small negative background, with value of order the image noise
level, to simulate a slight over subtraction.

\subsection{Overlapping Coadd Cells} \label{sec:sim:cells}

Coadds can be used for shear measurement \citep{ArmstrongCoadd}, with the
requirement that the coadded images have a continuous PSF in order to
facilitate accurate PSF modeling.  Images that have an edge in the coadd region
must be rejected.  Smaller coadds result in fewer images being rejected, but
very small coadds would complicate object identification and deblending.  For
LSST, we plan to use relatively small (250x250 pixels) overlapping (50 pixels)
coadded regions for shear measurement, which we call ``cells'' because they are
small subregions of the larger ``tract'' regions defined within the \dm\
framework.

Each cell definition has an extra boundary area that overlaps neighboring
cells.  This extra area provides a buffer that reduces measurement problems
that can occur at the image edge.  Objects with a measured center inside
overlaps are trimmed after the detection phase to avoid duplication.

We ran simulations with and without cells in order to control for issues
specific to the cell processing and trimming of objects to non overlapping
regions.


\subsection{Shear Patterns} \label{sec:sim:shears}

We implemented two different types of shear pattern.  In one we used a single
shear for all images, with constant magnitude and orientation.  The second type
was a constant magnitude shear, but with random orientation for each image.
These two scenarios result in equivalent shear recovery bias when objects are
not divided into cells on the sky (see \S \ref{sec:sim:cells}) but, as we will
see in \S \ref{sec:results:cells}, a small additional selection bias is
introduced for random shears when trimming the detected objects to the unique
cell regions.  In all cases the simulated shear had magnitude of \mshear.

\subsection{Noise Images} \label{sec:sim:noiseimages}

The \mcal\ procedure involves deconvolution by the PSF, shearing of the image,
and reconvolution by a round kernel.  This process alters the noise and
produces a spurious shear response \citep{SheldonMcal2017}.  We can cancel this
bias by adding a noise image, with properties that statistically match the real
background noise, processed with the orthogonal shear
\citep{SheldonMcal2017,mdet20}.

It is important that these noise images also contain the same correlated noise
as the real data. We followed the procedure outlined in \citet{BeckerMdetCoadd}
to generate the noise images. Namely, because the images are manipulated before
coaddition, for example interpolating saturated regions, bad columns, etc., we
passed the noise images through these manipulations as well. Furthermore,
warping will produce correlated noise, and thus the noise images are generated
with the same WCS as the real images, and the same warping is applied.

We stored this noise image in a copy of the image \calexp\ structure, but
replacing the real image data with the noise image.

\section{Image and PSF Coaddition} \label{sec:coadding}

For coaddition, we used the coadding strategy described in
\citet{BeckerMdetCoadd}. We reimplemented that strategy using \dm\ algorithms
for the various steps in the package \texttt{descwl\_coadd}. This work uses
version 0.3.0\footnote{\url{https://github.com/LSSTDESC/descwl_coadd}} of
\texttt{descwl\_coadd}.

For warping the images, we used the \texttt{Lanczos3} interpolant. The warped
images were combined with the \texttt{AccumulatorMeanStack} from
\texttt{lsst.meas.algorithms} sub package. We used an inverse variance
weighting, based on the median of the variance image plane.

Before warping, we interpolated the simulated artifacts in each image such as
cosmic rays, bad columns, and saturated pixels.  The interpolation and warping
modify the noise properties of the image.  Thus, for accurate \mcal\ noise
corrections (see \S \ref{sec:sim:noiseimages}) we must also run the noise image
through the same procedures.  Note stars were masked in the coadds, rather than
the input images (see \S \ref{sec:sim:satbleeds}).

Rather than use the \dm\ interpolation codes, we followed \citet{BeckerMdetCoadd},
performing all interpolation of artifacts and saturated regions using the
\texttt{CloughTocher2DInterpolator} from the scipy software
package\footnote{\url{https://docs.scipy.org/doc/scipy/reference/generated/scipy.interpolate.CloughTocher2DInterpolator.html}}.
We used this interpolation because the \dm\ cosmic ray interpolation available
at the time of writing is intertwined with the detection of cosmic rays, so
cannot be run on the noise image.

As shown in \citet{ArmstrongCoadd} and \citet{BeckerMdetCoadd}, creating a PSF
image for the coadd requires detailed tracking of the sub-pixel offsets of the
individual images in the coadd. For each input image, we generated an image of
the PSF at the location of the coadd center in the input image. The coadd image
center may fall at an arbitrary sub-pixel offset in the input image, and the
PSF was drawn with a matching offset.  We then warped these PSF images and
coadded them using the same weights as the image data. The off-centering and
warping is important so that the PSF includes the same small amount of smearing
present in the image interpolation. Not including this offset results in
percent level multiplicative shear biases \citep{ArmstrongCoadd}. Note \dm\
provides a PSF coadding code, but at the time of writing it did not offset the
PSFs before coadding, so was unsuitable for our purposes. The resulting coadd
and PSF coadd data were stored in a new \calexp\ data structure for use in
\mdet.

\begin{figure*}
    \begin{center}
    \includegraphics[width=0.65\textwidth]{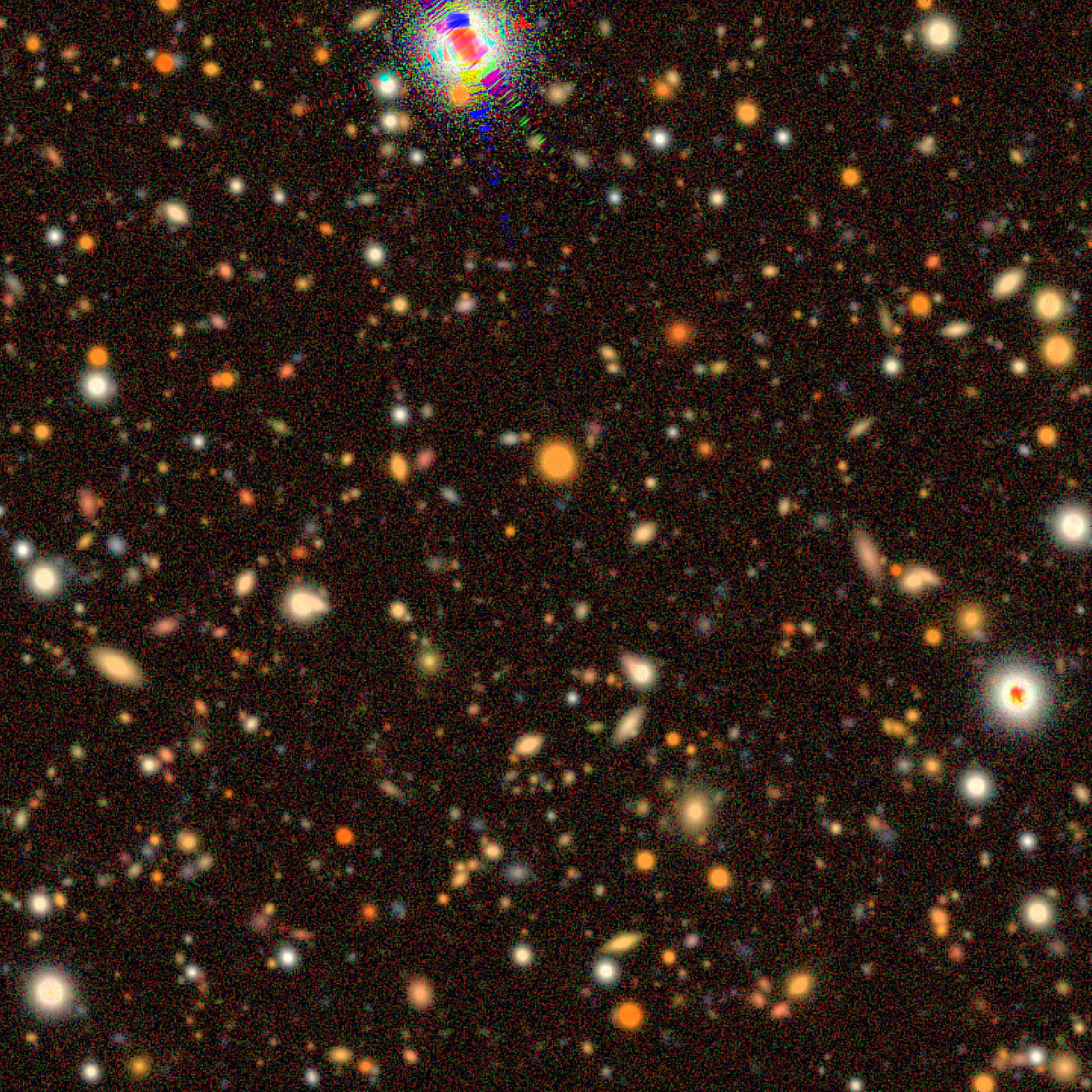}
    \caption{
        Example simulated image using a single image for each of the $r$, $i$,
        and $z$ bands, but with noise corresponding to the 10 year LSST coadd data.
        The images have rotations, dithers and image defects which are interpolated.
        The images were warped to a common coordinate system.
        Note the interpolated bleed trails around a bright star in the upper
        part of the image, appearing at different orientations due to the
        simulated camera rotations.
    } \label{fig:colorimage}
    \end{center}
\end{figure*}

\begin{figure}
    \includegraphics[width=\columnwidth]{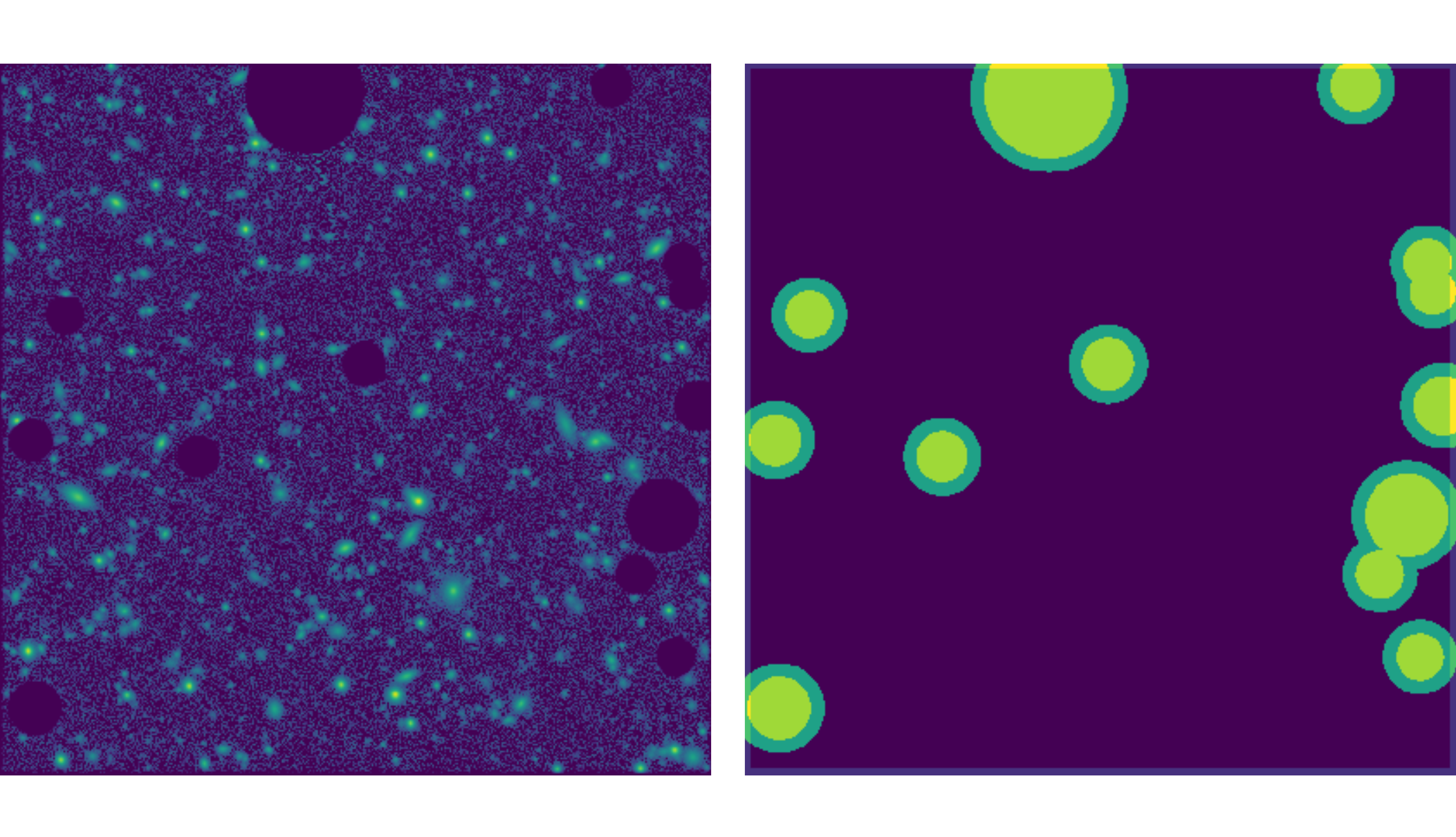}
    \caption{
        Example masked image corresponding to the $r$ band image used in figure
        \ref{fig:colorimage}, after star masking.  The left panel shows the
        masked image and the right shows the bitmask.  The circles represent
        the star masks applied to each coadd image before PSF deconvolution.
        The star masks are also marked in the bitmask to facilitate processing
        near stars.  The inner circle in the bitmask is the original mask size
        and the outer represents the expanded, apodized mask region.  In the
        original images, there are also pixels marked for cosmic rays and bad
        columns, but these are not retained in the coadd bitmask.
    } \label{fig:maskimage}
\end{figure}

In figure \ref{fig:colorimage} we show an example color image created from
warped $r$, $i$, and $z$ simulated images.  Figure \ref{fig:maskimage} shows
the corresponding masked images, along with the variance and bitmask planes.


\section{Metadetection} \label{sec:mdet}

For shear inference, we used the \mdet\ method presented in \cite{mdet20}.
\Mdet\ is an extension of the \mcal\ method
\citep{HuffMcal2017,SheldonMcal2017} to include the object identification
stage. We call this ``detection'', meaning specifically object identification
rather than detection of pixels with significant signal. \Mdet\ mitigates
biases due to the shear-dependent nature of the detection process in finite
resolution images, biases which are expected to be a few percent for LSST
\citep{mdet20}.

Briefly, for \mcal\ we assume that the true applied, two-component shear
$\boldsymbol{\gamma}$ is small, so that a measured ellipticity $\boldsymbol{e}$
of a galaxy is approximately linear in the shear:
\begin{eqnarray} \label{eq:response}
\boldsymbol{e} & \approx & \left.\boldsymbol{e}\right|_{\gamma=0} +
                           \left.\frac{\partial \boldsymbol{e}}{\partial\boldsymbol\gamma}\right|_{\gamma=0} \boldsymbol\gamma +
                           O(\boldsymbol\gamma^2)\nonumber\\
               & \equiv  & \left.\boldsymbol{e}\right|_{\gamma=0} +
                           \boldsymbol{R} \boldsymbol\gamma +
                           O(\boldsymbol\gamma^2)
\end{eqnarray}
Here, the matrix $\boldsymbol{R}$ represents the linear response of the
measurement to an applied shear. It can be written component wise as
\begin{equation}
R_{ij} = \frac{\partial e_i }{\partial \gamma_j } \biggr\rvert_{\gamma=0},
\end{equation}
with $i$ and $j$ taking all combinations of the two shear components.  For ellipticity
measurements we used weighted moments (see \S \ref{sec:mdet:meas}).

We use a finite difference to estimate $\boldsymbol{R}$.  The image is
deconvolved by the PSF, sheared, and reconvolved with a round kernel.  This
process is repeated for a small positive and negative shear, and a finite
difference response term is formed
\begin{equation}
R_{ij} \approx \frac{e_i^{+} - e_i^{-}}{\Delta\gamma_j}\ .
\end{equation}
where $e_i^{+}$ is measured on the positively sheared image and $e_i^{-}$ is
measured on the negatively sheared image.  The applied
shear can be any value much less than unity. We used counterfactual
shears $\gamma_j^{\pm} = \pm0.01$,
which gives $\Delta\gamma_j = 0.02$.

As mentioned in \S \ref{sec:sim:noiseimages}, the deconvolution, shear,
reconvolution process produces a spurious shear response that we correct by
adding a noise image run through the same process but with orthogonal shear.

The responses are too noisy to be applied to individual ellipticity measurements.
We instead average the shapes and responses in equation \ref{eq:response} to
recover a mean shear \vecg.  The mean response is
\begin{equation}
    \left< R_{ij} \right> = \left< \frac{\partial e_i }{\partial \gamma_j } \biggr\rvert_{\gamma=0} \right>,
\end{equation}
which for a finite difference estimate is
\begin{equation}
    \langle R_{ij}\rangle = \left< \frac{e_i^{+} - e_i^{-}}{\Delta\gamma_j} \right>.
\end{equation}
A mean shear can then be estimated using
\begin{equation} \label{eq:shearmeas}
    \langle \vecg \rangle \approx \langle \boldsymbol{R}\rangle^{-1}\langle\boldsymbol{e}\rangle.
\end{equation}
These averages may be calculated over the entire set of detected objects or
subsets of the data as needed, so long as the response can be determined with
sufficient precision.

However, this process is incomplete:  the detection phase also depends on the
shear.  We can incorporate detection by finding objects in the unsheared
and sheared images independently.  Moving the derivative outside of
the average, we find
\begin{equation}
    \left< R_{ij} \right> = \frac{\partial \left< e_i \right> }{\partial \gamma_j } \biggr\rvert_{\gamma=0},
\end{equation}
which for finite differences can be written
\begin{equation} \label{eq:fullR}
    \langle R_{ij}\rangle = \frac{\langle e_i^{+}\rangle - \langle e_i^{-}\rangle}{\Delta\gamma_j},
\end{equation}
where now the averages for $e_i^{+/-}$ are for object catalogs found on the
respective ${+/-}$ sheared image.

For \mdet, we used the python package \texttt{metadetect} version
0.8.2\footnote{\url{https://github.com/esheldon/metadetect}}.  In the
following subsections we will describe parts of this process in more detail.

\subsection{Creation of Sheared Images} \label{sec:mdet:sheared}

For this work we used \dm\ data structures to represent image data (see \S
\ref{sec:sim}).  We adapted the \mcal\ implementation for creating sheared
images from the \texttt{ngmix}
package\footnote{\url{https://github.com/esheldon/ngmix}} to use these data
structures.  The new code is in the \texttt{lsst} sub-package of
the \texttt{metadetect} repository.

\subsection{Detection and Deblending} \label{sec:mdet:detect}

We used the \dm\ peak finding algorithm for object detection
\citep{BoschHSC2017}.  We ran with the default settings at the time of writing,
which retains sources with signal-to-noise ratio \snr\ $\gtrsim 5$ as
calculated from a PSF template flux measurement.  Before detection we
determined the background in the coadd to correct for simulated background
estimation errors (see \S \ref{sec:sim:bgerr}).  We did not discard blended
objects.

For this work we did not use the deblender to remove the light of neighboring,
blended objects when measuring object properties (see \S \ref{sec:mdet:meas}).
We found that using the models to remove neighbor light resulted in shear
biases of order ten percent, which we suspect may be due to increased
non-linearity associated with this deblender.

We did limited testing with the alternate \texttt{Scarlet} deblender
\citep{MelchiorScarlet2018}\footnote{\url{https://github.com/pmelchior/scarlet}},
which is available to use with the \dm\ software stack. \texttt{Scarlet}
performed better in tests with pairs of galaxies at fixed
separation, but due to the high computational cost of the deblender, we did not
gather enough statistics to perform a comprehensive analysis.  We also did not
test \texttt{Scarlet} in more realistic scenes with many randomly places
galaxies.

As we will show in \S \ref{sec:results}, measuring shapes without deblending
the light of neighbors caused no detectable bias.  However, deblending may
improve the accuracy of other critical tasks such as inferring the redshift
distribution of the lensing source galaxies.  We will explore deblending more
in future work.

\subsection{Object Measurement} \label{sec:mdet:meas}

We used version v2.1.0 of the \ngmix\ package to measure weighted moments for
each detected object, using the \texttt{Gaussmom} class. The weight function
was a fixed, circular Gaussian $G(x, y)$ with FWHM=1.2 arcseconds. We chose
this FWHM to provide good precision for typical LSST seeing (FWHM=0.8
arcseconds), but this was not optimized.  We recorded the flux, S/N, second
moments, and ellipticity derived from the second moments:
\begin{eqnarray} \label{eq:moments}
    F &=& \sum G(x, y) I(x, y) \nonumber \\
    \sigma^2(F) &=& \sum G(x, y)^2 \sigma^2(I(x, y)) \nonumber \\
    T &=& \sum G(x, y) I(x, y) ~ (x^2 + y^2) \nonumber \\
    M_1 &=& \sum G(x, y) I(x, y) ~ (x^2 - y^2) \\
    M_2 &=& \sum G(x, y) I(x, y) ~ 2 x y \nonumber \\
    e_1 &=& M_1 / T \nonumber \\
    e_2 &=& M_2 / T \nonumber \\
    S/N &=& F / \sigma(F) \nonumber
\end{eqnarray}
where $I(x, y)$ is the intensity of the image, and $\sigma^2(I(x, y))$ is the
noise variance. The sums run over all pixels in a 48x48 postage stamp image
extracted around the object of interest.  The $x$ and $y$ coordinates are
relative to the center found during detection.

We measured the full covariance of the matrix of the moments, but for brevity
we have shown only the error on the flux $\sigma(F)$. Covariance elements can
all be generated from measurements of the general form $C_{i, j} = \sum G(x, y)^2
\sigma^2(I(x, y)) x^j y^j, \forall i, j \in [0, 2], i + j \le 2$.  We also
propagated the errors to derived quantities such as $e_1$ and $e_2$.  Note the
noise in the warped images is spatially correlated due to interpolation, but
only the zero lag variance was included in the the covariance calculations.

Note $T$ is an estimate of the observed object's size squared, not the
pre-PSF size.  We also record the moments of the coadd PSF, after
reconvolution, for use in object selection (see \S \ref{sec:results:full}).

\section{Running the Simulation and Metadetection} \label{sec:running}

We used the ``wrapper'' package \texttt{mdet-lsst-sim} version
0.3.4\footnote{\url{https://github.com/esheldon/mdet-lsst-sim}} to run the
simulation, coaddition and \mdet, and to organize, submit and collate large
runs on computing clusters.

In all cases we used a constant shear, with optional random shear orientations
for each simulated scene (see \S \ref{sec:sim:shears}).  We simulated each
scene twice, with equal but opposite shears, in order to implement the noise
canceling method of \cite{pujol2019}.  For randomized shear directions, the
shape measurements from different scenes were rotated into a common reference
frame before averaging to get a mean shear.

We estimated the uncertainty in the averaged shear measurements
$\sigma_{\gamma}$ using a jackknife technique \citep{LuptonStats1993}, with
chunks defined as a few tens to hundreds of scenes.  The jackknife technique
gives a significantly larger uncertainty than a naive error propagation, which
we attribute to additional sources of error associated with masking,
deblending, stellar contamination and variable response.

\section{Results} \label{sec:results}

In this section we present the results for various simulation and analysis
configurations.  In all simulations we used a shear of \mshear, which we expect
to result in a bias of a few parts in ten thousand due to higher order
shear effects \citep{SheldonMcal2017,mdet20}.

We identify our ``base'' simulation as one with fixed size, high S/N, round
galaxies on a grid layout, which provides a high signal-to-noise ratio for the
shear recovery.

We performed shear recovery tests starting with the base simulation and turning
on additional features successively in order to isolate the cause of potential
biases.  We found biases exceeding LSST requirements only in the case of
extreme spatial PSF variation combined with few simulated images contributing
to the coadd, so that the random variations were not sufficiently averaged down
(see \S \ref{sec:results:psfvar}).  We present results including all simulation
features in \S \ref{sec:results:cells}.

To characterize the bias, we assume a simple linear model \citep[see,
e.g.,][]{heymans2006} for estimating a bias in the recovered shear
\begin{equation} \label{eq:m}
\vecg = \vecc + (1 + \vecm) \vecgam
\end{equation}
where \vecg\ is the inferred shear, measured using equation \ref{eq:shearmeas},
\vecgam\ is the true shear, \vecc\ is the additive bias, and \vecm\ is the
multiplicative bias. We found \vecc\ was consistent with zero in all tests, so
in what follows we only report \vecm.  Furthermore, we only report
the scalar $m$, even in the case of random shear orientations, because
the ellipticities are rotated to a common frame before averaging (see \S
\ref{sec:running}).

\subsection{LSST Requirements} \label{sec:lsstreq}

We adopt the requirements from \cite{SRD}, who derived a limit on the
redshift-dependent multiplicative bias, for each of $N_{\mathrm{tomo}}$
tomographic bins, as $|m_{\mathrm{tomo}}| < 0.003$.  Because our simulation
does not contain redshifts, we will essentially constrain an overall bias
across all redshifts.  The bias will be partly correlated across bins, so we
expect the overall requirement to be lower than $m_{\mathrm{tomo}}$, but higher
than $m_{\mathrm{tomo}}/\sqrt{N_{\mathrm{tomo}}}$.  In what follows, we adopt
$|m| < $\mthresh\ as an intermediate value, but we want to emphasize that this
value is somewhat arbitrary.

\subsection{Baseline Results for Identical Galaxies on a Grid} \label{sec:results:base}

In order to establish a baseline for the bias due to higher order shear
effects, we ran a simulation with identical, round exponential, S/N $= 10,000$,
half light radius 0.5 arcsecond galaxies placed in the grid layout with a 9.5
arcsecond spacing, for a density of 40 per square arcminute (see sections
\ref{sec:sim:galaxies} and \ref{sec:sim:layouts}), with image dithers and
rotations.  In order to increase the computationally efficiency, we used only
the $i$ band with a single image warped to the coadd frame, rather than many
images warped and coadded.  We used a fixed, circular Moffat
PSF\citep{Moffat1969} with FWHM=0.8.  We applied a cut $S/N > 10$ to remove
spurious detections.  We expect a bias $m$ of a few parts in ten thousand for
this sim \citep{SheldonMcal2017}, and indeed we found a bias of $4.2\times
10^{-4} < m < 4.3\times 10^{-4}$ (99.7\%~confidence), consistent with our
expectations.

We reran this relatively fast simulation test after all major code updates as a
kind of extended unit test to expose bugs and regressions.

\subsection{PSF Variation} \label{sec:results:psfvar}

In order to measure the shear response, the \mdet\ algorithm creates
artificially sheared versions of each coadd image (see \S \ref{sec:mdet}).
This process involves deconvolving the image by the PSF (see \S
\ref{sec:mdet}).  For a spatially varying PSF, this deconvolution would require a
spatially varying kernel.  However, due to the large number of images used to
make the coadd, and camera rotations and dithers which will partially randomize
any static PSF patterns, we expect the spatial variation in the final image
to be significantly reduced.  Thus the single coadded PSF that we generate at
the coadd center (see \S \ref{sec:sim:psfs}) may be sufficient.

We ran the same grid simulation from \S \ref{sec:results:base} with the
spatially varying PSF presented in \S \ref{sec:sim:psfs}. We used the $r, i, z$
bands but with a single warped image per band, with rotations and dithers,
rather than many images warped and coadded.  We did not see any increase in the
multiplicative bias.   Our interpretation is that the PSF variation in the
coadd is small enough that our single coadded PSF image is sufficient, even
with a single image per band.

We also ran with the same simulation configuration, but with ten times the
expected PSF variation. In this case, for a single image per band, we found a
bias $m \sim 0.0016$.  However, after increasing the number of simulated images
per band to ten, warped and coadded (but with the same final coadd noise
level), the bias reduced to that expected from higher order shear effects.  We
interpret this to mean that more coadded images are required to average down
very large PSF variations.

As argued in \citet{mdet20}, since LSST will take hundreds of images in each
band \citep{IvezicLSST2008}, we should not expect random PSF variations to be a
significant source of bias.  However, note that our PSFs included only random
patterns, not {\em complex static} PSF patterns due to the telescope and
camera. Such patterns, while subdominant for LSST PSFs, would reduce more
slowly when combining images, especially for poorly chosen dither
patterns. We will test more complex PSF patterns in a future work.

For the remaining tests in this paper that employ a spatially variable PSF, the
expected level of variation was used.

\subsection{Results with Additional Simulation Features} \label{sec:results:more}

We ran the simulation with realistic galaxies, random layout, image
artifacts, and background subtraction errors, as listed in \S \ref{sec:sim}.

For the realistic galaxy configuration, we adopted a basic set of cuts for use
in quoting the results
\begin{eqnarray} \label{eq:basiccuts}
    \mathrm{S/N} & > & 12.5 \\
    T/T_{PSF} & > & 1.2
\end{eqnarray}
where $T$ is the Gaussian weighted size squared (see equation \ref{eq:moments})
of the object, as defined in equation \ref{eq:moments}, and $T_{PSF}$
is the corresponding value for the PSF, both measured after the \mcal\
reconvolution step. The lower \snr\ cut is motivated by the noise study in \S
\ref{sec:results:sdensnoise} in which a cut at 10 or 12.5 resulted in similar
shear measurement uncertainties. In all the following tests, we performed the
measurements with a range of cuts, and a lower bound of $S/N > 10$ always gave
results consistent with the default cut.

The lower cut on $T/T_{PSF}$ represents a simple attempt to remove stars from
the sample.  For \mcal, including stars in the sample does not result in a
biased shear when the PSF is accurately modeled \citep{SheldonMcal2017}.  With
\mdet\ the stars may result in a small bias because their positions are not
moved by the true shear, but {\it are} moved by the artificial shear. However,
the bias caused by this counterfactual movement is small \citep{mdet20}.
Nevertheless, it may be desirable to remove stars for other reasons, such as to
reduce noise and avoid confusion when determining the redshift distribution, so
we include a crude star removal in our tests.

We again turned on each feature successively in order to isolate possible
biases.  In no case did we find bias in the shear recovery larger than our
accuracy goals (see \S \ref{sec:lsstreq}).  More details are given in the
following sub-sections.

\subsubsection{Results with Varied Selections on Object Properties} \label{sec:results:select}

We varied the selections on $T/T_{PSF}$ and \snr\ and
repeated the shear recovery analysis.  The results are shown in figure
\ref{fig:trends}, for both constant and spatially varying PSFs (see \S
\ref{sec:sim:psfs}).  We did not detect any additional bias.
Counterintuitively, higher \snr\ thresholds resulted in lower noise for this
test.  Normally removing measurements would result in a more noisy mean, but
the noise cancellation procedure works better for high \snr\ objects that are
more often detected in both simulations.  We will explore the shear sensitivity
without noise cancellation in \S \ref{sec:results:sdensnoise}.

%
%

\input{trends.tex}

\subsection{Results with Overlapping Coadd Cells} \label{sec:results:cells}

We ran a modification of the base simulation as described in \S
\ref{sec:results:base} with a randomized position layout and overlapping cells
as described in \S \ref{sec:sim:cells}.  Because the cells overlap, the catalog
created for each coadd must be trimmed to a unique region.  Shear changes the
locations of objects, so this trimming introduces a shear dependent selection
effect.  For \mdet\ to correct for this selection, it is important that the
full scene be sheared, so that the object positions change after application of
the counterfactual shear.

In the case where the true simulated shear aligns with the counterfactual shear
used to create the \mcal\ images, we found that \mdet\ provided an accurate
calibration.  However, when using randomized shear orientations (see \S
\ref{sec:sim:shears}), we found the multiplicative bias $m$ was lower than the
expected bias due to higher order shear effects by about $-0.0004$.  Our
interpretation is that the artificial shearing slightly over-predicts the
selection effect when it is not perfectly aligned with the true shear.  This
bias is smaller than LSST requirements $|m|\lesssim~$\mthresh.  Nevertheless, we
speculate that using artificial shears with random orientations could mitigate
this effect.  We may explore this more in a future work.

\subsection{Results with Stars and Bleed Trails} \label{sec:results:full}

Our ``full'' simulation configuration included stars and bleed trails as
described in \S \ref{sec:sim:stars} and \ref{sec:sim:satbleeds}, in addition to
the features tested in the previous sections.   We discuss aspects of the
measurements and results below. For these tests we simluated a total of \fullarea\
square degrees with this configuration.

\subsubsection{Weighting}

For the full simulation configuration we found it beneficial to apply weights
in the shear average.  We used a two part weight function
\begin{equation}
    W(\sigma_e, e) = W_\sigma (\sigma_e) \times W_e(e)
\end{equation}
with $W_\sigma(\sigma_e)$ based on the object's ellipticity noise and $W_e(e)$
designed to down-weight a particular set of spurious objects.

The $W_\sigma(\sigma_e)$ is an approximate inverse variance weighting
\begin{equation} \label{eq:wsigma}
    W_\sigma = \frac{1}{\sigma_{SN}^2 + \sigma_e^2},
\end{equation}
where $\sigma^2_{SN}$ is the intrinsic ellipticity variance of the population,
before response correction, and $\sigma_e$ is the estimated uncertainty of the
object's ellipticity from shot noise (see \S \ref{sec:mdet:meas}).  We find
$\sigma_{SN}/R \sim 0.29$ for high \snr\ objects in our catalog.  The average
response is about 0.24 for this sample, so the pre-response value used
in the weight is 0.07.

In the presence of stars we found an unexpected population of high ellipticity
objects.  These objects were usually spurious detections or blends of faint
objects with bright stars, found just outside the star masking
radius\footnote{Masaya Yamamoto, private communication} (see \S
\ref{sec:sim:satbleeds}).  Including these objects did not produce a bias in
the inferred shear, but did increase the noise.  Due to a lack of foresight, we
did not keep enough information in the output files to exclude these objects
from our catalogs in post-processing, and rerunning would have taken
prohibitively long due to limited computing resources.  As a temporary solution
we included an additional ellipticity dependent weight function \citep{ba14}
\begin{equation}
    W_e(e) = \left[1 - e^2\right]^2 \mathrm{exp}\left[ -e^2/2 \sigma_w^2\right]
\end{equation}
where $e$ is the measured ellipticity magnitude before response
correction $e = \sqrt{e_1^2 + e_2^2}$.  We took
$\sigma_w = 0.3$, which downweights very high ellipticity objects.  We found
that using this weight function reduced the uncertainty of the shear recovery
by 13\%.  Note, in this case, the improvement in the uncertainty in the mean
shear is not due primarily to a reduction in shape noise for real objects, but
due to downweighting spurious detections.  In future work it will be important
to use sufficient star masking, and identify problematic objects.

\subsubsection{Dependence of Shear Bias on Stellar Density Cut} \label{sec:results:sdens}

In figure \ref{fig:stardens} we show the multiplicative bias as a function of
maximum stellar density and various cuts on the measured signal-to-noise ratio,
in addition to a common cut on size ratio $T/T_{PSF} > 1.2$.  In no case did we
detect a bias larger than our accuracy goals (see \S \ref{sec:lsstreq}).

\input{stardens-bias.tex}

\subsubsection{Dependence of Shear Sensitivity on Object Selections}
\label{sec:results:sdensnoise}

We explored the dependence of the shear sensitivity on stellar density and
\snr.  In order to get an accurate measure of the relative uncertainty, we did
not cancel noise using the results from paired images.  The noise cancellation
is most effective for higher \snr\ objects that are more likely to be detected
in both of the paired images, so using it would result in artificially smaller
uncertainties for higher \snr\ thresholds.

In figure \ref{fig:stardenserror} we show the uncertainty in the recovered
shear as a function of maximum allowed stellar density and \snr\ cut, relative
to the cuts that provide the lowest noise ($S/N > 12.5$ and $T/T_{PSF} > 1.2$).

We find the uncertainty decreases with the maximum allowed stellar density. The
dependence on stellar density is fairly flat at high stellar density; including
higher stellar density areas reduces the noise, but with limited benefit beyond
a density of about 40 stars per square arcminute.  This is expected because a
relatively small fraction of the simulated area has high stellar density.

\input{stardens-nocancel-error.tex}

We found that including detections with \snr\ as low as 10 does not reduce the
shear uncertainty compared to cutting at 12.5; both cuts were basically
equivalent.  This may be because lower \snr\ objects have relatively small
response, which is not accounted for in the weighting.  An optimized weight
based on the expected response for each object may provide a better sensitivity
\citep{GattiY3Shear2021}.  Note also that these results will vary significantly
with the measurement type, for example the specific weight function used to
measure moments, or the model that is fitted.

\subsubsection{Discussion of Practical Stellar Density Cuts}

Although we found no significant increase in shear bias when including images
with higher stellar density, nor did we see degradation in the statistical
uncertainty, it may be beneficial to exclude them to reduce the impact of
sources of error that correlate with stellar density.  For example, high
interstellar extinction and high stellar contamination in the shear sample
could cause biases in the estimated redshift distribution.  The stellar density
cut could be tuned to provide a beneficial trade off between accuracy and shear
sensitivity.

\section{Discussion} \label{sec:summary}

We have shown that \mdet\ is robust to the types of image features we expect
the shear code to handle, namely those that do not require further
characterization or calibration of the input data.

The noise propagation in particular requires careful handling.  Problematic
features, such as saturation and bright stars, must be sufficiently masked and
interpolated, and the images will most likely be coadded for efficiency. Both
of these steps significantly modify the noise, introducing correlations that
can bias the shear recovery.  We find that
the shear can be accurately recovered if representative noise images are passed
through the same image manipulations and used to restore sufficient symmetry to
the noise following the procedures in \cite{SheldonMcal2017,BeckerMdetCoadd}.

We expect further simulation work will be critical moving forward.  We plan to
proceed on three separate, complimentary tracks.  The first track is to generate simulations to
jointly calibrate the redshift dependent shear bias and redshift distribution
$N(z)$ of the sources, following the pioneering work of \cite{MacCrann2022} and
\cite{LiNofz2022}.  Unlike the simulations used here, the calibration
simulations may need to reproduce specific characteristics of a representative
sample of coadd cells, including the correlated noise, masking, PSF etc.  We
expect this joint calibration to be necessary even if the shear measurement
method itself is unbiased.  An alternative to a pure simulation approach may to
inject fake sources directly into the images
\citep{SuchytaBalrog2016,EverettBalrog2022}, which more naturally captures the
characteristics of the original images. However, this approach may be computationally
less efficient. In order to avoid excessive perturbations to the images,
the sources must be injected with low density, requiring
more images to be processed to provide the same precision.

The second simulation track is to test the effect of specific errors in the
data characterization.  For example, PSF determination errors, background
estimation errors (more complex than we tested in this work), incorrect image
noise determination, or other kinds of miscalibration.  It may be that
relatively small, targeted simulations are sufficient to estimate the impact of
these errors on shear measurement.

The third track is to adjust the procedures described here in order to optimize
the overall precision of the shear measurement. Possible research directions in
this vein include finding optimal object measurements and galaxy selections,
and generalizing \mdet\ to use the more precise calibrations
provided by the newly developed deep-field \mcal\ \citep{dfmcal22}.

\section*{Acknowledgments}

This paper has undergone internal review in the LSST Dark Energy Science
Collaboration.  We thank the internal reviewers, Xiangchong Li, Arun Kannawadi,
and Tianqing Zhang.  Thanks to Javier Sanchez for providing the star and
stellar density map for DC2, and Jim Chiang for providing example images
containing simulated stars with bleed trails.  We thank the excellent computing
staffs of the RHIC Atlas Computing Facility at Brookhaven National Laboratory
and the Scientific Computing Services team at the SLAC National Accelerator
Laboratory for their support.  ES is supported by DOE grant DE-AC02-98CH10886,
and MRB is supported by DOE grant DE-AC02-06CH11357. RA is supported by the US
Department of Energy Cosmic Frontier program, grant DE-SC0010118.  MJ
acknowledges support from the LSST Dark Energy Science Collaboration.

Author contributions to this work are as follows:  E. Sheldon and M. Becker
co-developed the measurement and simulation codes. E. Sheldon developed new
warping and coaddition code using existing \dm\ methods based on an initial
implementation from R. Armstrong. E. Sheldon wrote most of the text and
performed the shear recovery runs and analysis on compute clusters.  M. Becker
wrote a significant portion of the text and performed independent validation of
many of the methods.  M. Jarvis provided detailed contributions for optimizing
the object drawing with \galsim.

\input{descack}

\bibliographystyle{aasjournal}
\bibliography{references}

\end{document}

%% file: trends.tex
\begin{figure}
    \centering
\begin{tikzpicture}
    \begin{groupplot}[
        group style={
            group size=1 by 2,
        },
        width=\columnwidth,
        height=0.7\columnwidth,
        minor tick num=3,
        axis on top,
        legend style={
            legend columns=2,
            draw=none,
        },
    ]

    %
    %
    \nextgroupplot[
        xlabel={Minimum S/N},
        ylabel={$m/10^{-3}$},
        xmin=9,xmax=21,
        ymin=-2.5,ymax=3.4,
    ]

	\addplot+[dashdotted,sand5,no markers,thick] coordinates {
        (9.0,0.4)
        (21.0,0.4)
    };
    \addlegendentry{Nonlinear Shear}

	\addplot+[name path=A,bisque,no markers,forget plot] coordinates {
        (9.0,-2)
        (21.0,-2)
    };
    \addplot+[name path=B,bisque,no markers,forget plot] coordinates {
        (9.0,2)
        (21.0,2)
    };

    \addplot+[bisque,opacity=0.5] fill between[of=A and B];
    \addlegendentry{Requirement}

	\addplot+[name path=mlowvar,desert3,no markers,solid,opacity=0.5,forget plot] coordinates {
        (10.0,0.219)
        (12.5,0.078)
        (15.0,0.213)
        (20.0,0.257)
    };
    \addplot+[name path=mhighvar,desert3,no markers,solid,opacity=0.5,forget plot] coordinates {
        (10.0,1.215)
        (12.5,0.869)
        (15.0,0.973)
        (20.0,0.892)
    };
    \addplot+[color=desert3,opacity=0.5] fill between[of=mlowvar and mhighvar];
    \addlegendentry{LSST 10 year PSF}

	\addplot+[name path=mlowfix,bsand4,no markers,solid,opacity=0.5,forget plot] coordinates {
        (10.0,0.176295)
        (12.5,-0.00826672)
        (15.0,0.165007)
        (20.0,0.111993)
    };
    \addplot+[name path=mhighfix,bsand4,no markers,solid,opacity=0.5,forget plot] coordinates {
        (10.0,1.13744)
        (12.5,0.789173)
        (15.0,0.915298)
        (20.0,0.765478)
    };
    \addplot+[color=bsand4,opacity=0.5] fill between[of=mlowfix and mhighfix];
    \addlegendentry{Constant PSF}

	\addplot+[solid,black!80,no markers,forget plot] coordinates {
        (9.0,0.0)
        (21.0,0.0)
    };

    %
    %
    \nextgroupplot[
        xlabel={Minimum $T/T_{PSF}$},
        ylabel={$m/10^{-3}$},
        xmin=1.15,xmax=1.55,
        ymin=-2.5,ymax=3.4,
    ]

	\addplot+[dashdotted,sand5,no markers,thick] coordinates {
        (1.15,0.4)
        (1.55,0.4)
    };

	\addplot+[name path=A,bisque,no markers,forget plot] coordinates {
        (1.15,-2)
        (1.55,-2)
    };
    \addplot+[name path=B,bisque,no markers,forget plot] coordinates {
        (1.15,2)
        (1.55,2)
    };

    \addplot+[bisque,opacity=0.5] fill between[of=A and B];

	\addplot+[name path=mlowvar,desert3,no markers,solid,opacity=0.5,forget plot] coordinates {
        (1.2, 0.211992)
        (1.3, 0.151222)
        (1.4, -0.0532513)
        (1.5, -0.346328)

    };
    \addplot+[name path=mhighvar,desert3,no markers,solid,opacity=0.5,forget plot] coordinates {
        (1.2, 0.967327)
        (1.3, 1.04341)
        (1.4, 0.9606359999999999)
        (1.5, 0.968678)

    };
    \addplot+[color=desert3,opacity=0.5] fill between[of=mlowvar and mhighvar];

    \addplot+[name path=Tratmlowfix,bsand4,no markers,solid,opacity=0.5,forget plot] coordinates {
        (1.2, 0.165007)
        (1.3, -0.0856819)
        (1.4, -0.114263)
        (1.5, -0.370727)
    };
    \addplot+[name path=Tratmhighfix,bsand4,no markers,solid,opacity=0.5,forget plot] coordinates {
        (1.2, 0.9152980000000001)
        (1.3, 0.799206)
        (1.4, 0.858479)
        (1.5, 0.9879929999999999)
    };
    \addplot+[color=bsand4,opacity=0.5] fill between[of=Tratmlowfix and Tratmhighfix];

	\addplot+[solid,black!80,no markers,forget plot] coordinates {
        (1.15,0.0)
        (1.55,0.0)
    };

    \end{groupplot}
\end{tikzpicture}

    \caption{
        Multiplicative shear bias $m$ for a simulation with a realistic galaxy
        sample, random galaxy layout, image artifacts and background
        subtraction errors.  The top panel shows the bias as a function of
        \snr\ and the bottom panel shows the bias as a function of the ratio of
        the size squared of the object to that of the PSF, \Tratio.  The shaded
        areas show the 99.7\%~confidence bands for the constant PSF (darker)
        and spatially variable PSF (lighter).  The dot-dashed line represents
        the approximate expected bias due to higher order shear effects, and
        the broader shaded band represents our nominal accuracy goal.   In all
        cases the results are consistent with
        the expected bias.
    }
    \label{fig:trends}

\end{figure}
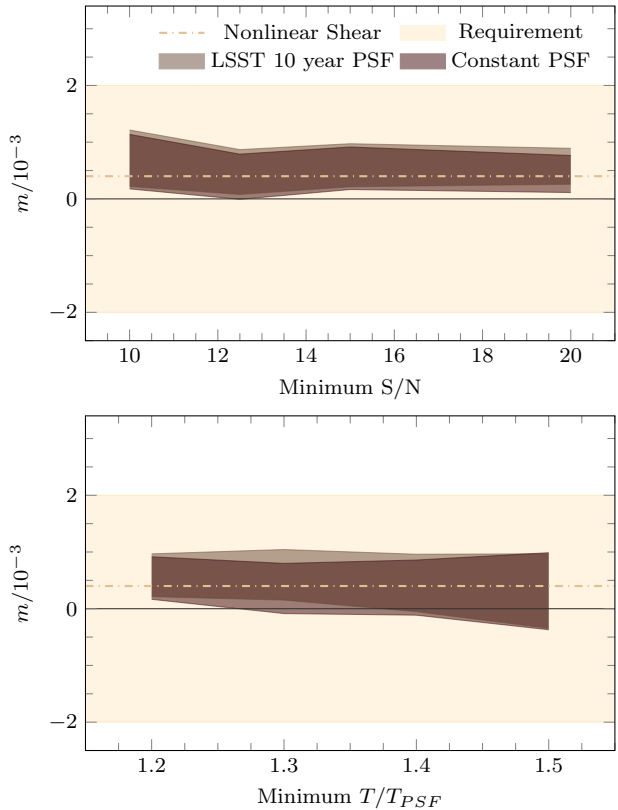

%% file: stardens-bias.tex
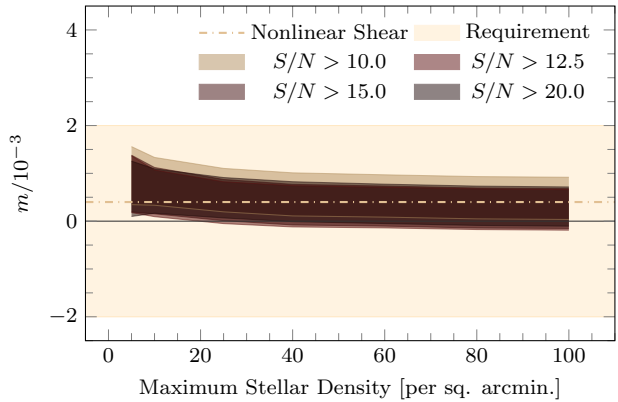
\begin{figure}
    \centering
\begin{tikzpicture}
    \begin{axis} [
        xlabel={Maximum Stellar Density [per sq. arcmin.]},
        ylabel={$m/10^{-3}$},
        xmin=-5,xmax=110,
        ymin=-2.5,ymax=4.5,
        width=\columnwidth,
        height=0.7\columnwidth,
        minor tick num=3,
        axis on top,
        legend style={
            legend columns=2,
            draw=none,
        }
    ]

	\addplot+[dashdotted,sand5,no markers,thick] coordinates {
        (-5,0.4)
        (110.0,0.4)
    };
    \addlegendentry{Nonlinear Shear}

	\addplot+[name path=A,bisque,no markers,forget plot] coordinates {
        (-5.0,-2)
        (110,-2)
    };
    \addplot+[name path=B,bisque,no markers,forget plot] coordinates {
        (-5,2)
        (110,2)
    };

    \addplot+[bisque,opacity=0.5] fill between[of=A and B];
    \addlegendentry{Requirement}

    \addplot+[
        name path=mlowten,bsand1,no markers,solid,opacity=0.5,forget plot
    ] table[x index=0, y index=1] {code/s2n-10.0-Tratio-1.2.txt};
    \addplot+[
        name path=mhighten,bsand1,no markers,solid,opacity=0.5,forget plot
    ] table[x index=0, y index=2] {code/s2n-10.0-Tratio-1.2.txt};
    \addplot+[bsand1,opacity=0.5,solid] fill between[of=mlowten and mhighten];
    \addlegendentry{$S/N > 10.0$}

    \addplot+[
        name path=mlowtwel,bsand3,no markers,solid,opacity=0.5,forget plot
    ] table[x index=0, y index=1] {code/s2n-12.5-Tratio-1.2.txt};
    \addplot+[
        name path=mhightwel,bsand3,no markers,solid,opacity=0.5,forget plot
    ] table[x index=0, y index=2] {code/s2n-12.5-Tratio-1.2.txt};
    \addplot+[bsand3,opacity=0.5,solid] fill between[of=mlowtwel and mhightwel];
    \addlegendentry{$S/N > 12.5$}

    \addplot+[
        name path=mlowfif,bsand4,no markers,solid,opacity=0.5,forget plot
    ] table[x index=0, y index=1] {code/s2n-15.0-Tratio-1.2.txt};
    \addplot+[
        name path=mhighfif,bsand4,no markers,solid,opacity=0.5,forget plot
    ] table[x index=0, y index=2] {code/s2n-15.0-Tratio-1.2.txt};
    \addplot+[bsand4,opacity=0.5,solid] fill between[of=mlowfif and mhighfif];
    \addlegendentry{$S/N > 15.0$}

    \addplot+[
        name path=mlowtwen,bsand5,no markers,solid,opacity=0.5,forget plot
    ] table[x index=0, y index=1] {code/s2n-20.0-Tratio-1.2.txt};
    \addplot+[
        name path=mhightwen,bsand5,no markers,solid,opacity=0.5,forget plot
    ] table[x index=0, y index=2] {code/s2n-20.0-Tratio-1.2.txt};
    \addplot+[bsand5,opacity=0.5,solid] fill between[of=mlowtwen and mhightwen];
    \addlegendentry{$S/N > 20.0$}

	\addplot+[solid,black!80,no markers,forget plot] coordinates {
        (-5, 0.0)
        (110, 0.0)
    };

    \end{axis}
\end{tikzpicture}

	\caption{
        Multiplicative bias $m$ as a function of maximum stellar density cut
        for various cuts on \snr. The shaded areas show the 99.7\%~confidence bands
        for the various cuts.  The dot-dashed line represents the
        approximate expected bias due to higher order shear effects, and the
        broader shaded band represent our nominal accuracy goal.  In no case did we
        detect any bias larger than expectations.
    } \label{fig:stardens}

\end{figure}

%% file: stardens-nocancel-error.tex
\begin{figure}
    \centering
\begin{tikzpicture}
    \begin{axis}[
        xlabel={Maximum Stellar Density [per sq. arcmin.]},
        ylabel={$\sigma(\gamma)/\sigma_{min}(\gamma)$},
        xmin=-5,xmax=110,
        ymin=0.95,ymax=1.65,
        width=\columnwidth,
        height=0.7\columnwidth,
        minor tick num=3,
        axis on top,
        legend style={
            draw=none,
        },
    ]

    \addplot+[
        teal!60!black,
        mark=square*,
        mark options={fill=teal!40},
    ] table[x index=0, y index=1] {code/s2n-20.0-Tratio-1.2-err.txt};
    \addlegendentry{$S/N > 20.0$}

    \addplot+[
        red!60!black,
        mark=triangle*,
        mark options={fill=red!40},
        solid,
    ] table[x index=0, y index=1] {code/s2n-15.0-tratio-1.2-err.txt};
    \addlegendentry{$S/N > 15.0$}

    \addplot+[
        PineGreen!60!black,
        mark=diamond*,
        mark options={fill=PineGreen!40},
        solid,
    ] table[x index=0, y index=1] {code/s2n-12.5-Tratio-1.2-err.txt};
    \addlegendentry{$S/N > 12.5$}

    \addplot+[
        brown!60!black,
        mark=otimes*,
        mark options={fill=brown!40},
        solid,
    ] table[x index=0, y index=1] {code/s2n-10.0-Tratio-1.2-err.txt};
    \addlegendentry{$S/N > 10.0$}

	\addplot+[solid,black!80,no markers,forget plot] coordinates {
        (-5, 1.0)
        (110, 1.0)
    };

    \end{axis}
\end{tikzpicture}

    \caption{
        Shear measurement sensitivity as a function of maximum stellar density
        and \snr, relative to the cuts that give the best sensitivity.  We find
        that less restrictive cuts on stellar density result in better
        sensitivity, but with marginal improvement beyond a density of about 40
        per square arcminute due to lack of area with high stellar density.
        Keeping objects with \snr\ as low as 10 did not give an improved
        sensitivity compared to a cut at 12.5.  Note that these results may
        differ for alternative ellipticity measurements or weighting schemes.
    } \label{fig:stardenserror}

\end{figure}
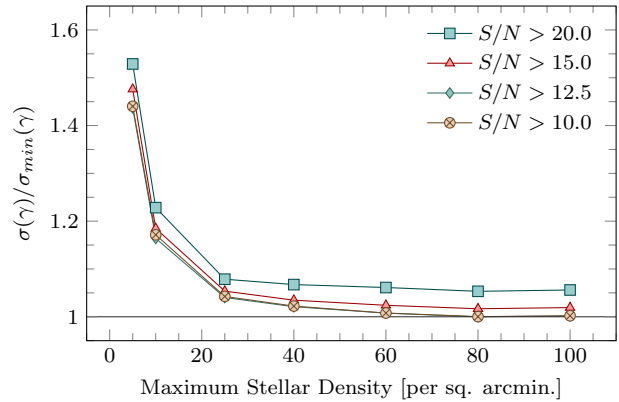

%% file: descack.tex
DESC acknowledges ongoing support from the IN2P3 (France), the STFC 
(United Kingdom), and the DOE, NSF, and LSST Corporation (United States).  
DESC uses resources of the IN2P3 Computing Center 
(CC-IN2P3--Lyon/Villeurbanne - France) funded by the Centre National de la
Recherche Scientifique; the National Energy Research Scientific Computing
Center, a DOE Office of Science User Facility supported under Contract 
No.\ DE-AC02-05CH11231; STFC DiRAC HPC Facilities, funded by UK BEIS National 
E-infrastructure capital grants; and the UK particle physics grid, supported
by the GridPP Collaboration.  This work was performed in part under DOE 
Contract DE-AC02-76SF00515.

%% file: mdetlsst.bbl
\begin{thebibliography}{}
\expandafter\ifx\csname natexlab\endcsname\relax\def\natexlab#1{#1}\fi
\providecommand{\url}[1]{\href{#1}{#1}}
\providecommand{\dodoi}[1]{doi:~\href{http://doi.org/#1}{\nolinkurl{#1}}}
\providecommand{\doeprint}[1]{\href{http://ascl.net/#1}{\nolinkurl{http://ascl.net/#1}}}
\providecommand{\doarXiv}[1]{\href{https://arxiv.org/abs/#1}{\nolinkurl{https://arxiv.org/abs/#1}}}

\bibitem[{Abolfathi {et~al.}(2021)}]{DC2Abolfathi2021}
Abolfathi, B., {et~al.} 2021, The Astrophysical Journal Supplement Series, 253,
  31, \dodoi{10.3847/1538-4365/abd62c}

\bibitem[{{Akeson} {et~al.}(2019)}]{AkesonRoman2019}
{Akeson}, R., {et~al.} 2019, arXiv e-prints, arXiv:1902.05569,
  \dodoi{10.48550/arXiv.1902.05569}

\bibitem[{{Armstrong, R., et al.}(in prep.)}]{ArmstrongCoadd}
{Armstrong, R., et al.} in prep., in prep

\bibitem[{{Becker} {et~al.}(in prep.){Becker}, {Sheldon}, \&
  {Jarvis}}]{BeckerMdetCoadd}
{Becker}, M.~R., {Sheldon}, E.~S., \& {Jarvis}, M. in prep., in prep.

\bibitem[{{Bernstein} \& {Armstrong}(2014)}]{ba14}
{Bernstein}, G.~M., \& {Armstrong}, R. 2014, \mnras, 438, 1880,
  \dodoi{10.1093/mnras/stt2326}

\bibitem[{{Bernstein} {et~al.}(2016){Bernstein}, {Armstrong}, {Krawiec}, \&
  {March}}]{BernBFD2016}
{Bernstein}, G.~M., {Armstrong}, R., {Krawiec}, C., \& {March}, M.~C. 2016,
  \mnras, 459, 4467, \dodoi{10.1093/mnras/stw879}

\bibitem[{Bosch {et~al.}(2017)}]{BoschHSC2017}
Bosch, J., {et~al.} 2017, Publications of the Astronomical Society of Japan,
  70, \dodoi{10.1093/pasj/psx080}

\bibitem[{{Bosch} {et~al.}(2019){Bosch}, {AlSayyad}, {Armstrong}, {Bellm},
  {Chiang}, {Eggl}, {Findeisen}, {Fisher-Levine}, {Guy}, {Guyonnet},
  {Ivezi{\'c}}, {Jenness}, {Kov{\'a}cs}, {Krughoff}, {Lupton}, {Lust},
  {MacArthur}, {Meyers}, {Moolekamp}, {Morrison}, {Morton}, {O'Mullane},
  {Parejko}, {Plazas}, {Price}, {Rawls}, {Reed}, {Schellart}, {Slater},
  {Sullivan}, {Swinbank}, {Taranu}, {Waters}, \& {Wood-Vasey}}]{BoschLSST2019}
{Bosch}, J., {AlSayyad}, Y., {Armstrong}, R., {et~al.} 2019, in Astronomical
  Society of the Pacific Conference Series, Vol. 523, Astronomical Data
  Analysis Software and Systems XXVII, ed. P.~J. {Teuben}, M.~W. {Pound}, B.~A.
  {Thomas}, \& E.~M. {Warner}, 521.
\newblock \doarXiv{1812.03248}

\bibitem[{{Everett} {et~al.}(2022)}]{EverettBalrog2022}
{Everett}, S., {et~al.} 2022, \apjs, 258, 15, \dodoi{10.3847/1538-4365/ac26c1}

\bibitem[{Gatti {et~al.}(2021)}]{GattiY3Shear2021}
Gatti, M., {et~al.} 2021, Monthly Notices of the Royal Astronomical Society,
  504, 4312, \dodoi{10.1093/mnras/stab918}

\bibitem[{Heymans {et~al.}(2012)Heymans, Kitching, Rowe, Hoekstra, Miller,
  Erben, \& Van~Waerbeke}]{heymans2012}
Heymans, C., Kitching, T., Rowe, B., {et~al.} 2012, Monthly Notices of the
  Royal Astronomical Society, 421, 381,
  \dodoi{10.1111/j.1365-2966.2011.20312.x}

\bibitem[{{Heymans} {et~al.}(2006){Heymans}, {Van Waerbeke}, {Bacon}, {Berge},
  {Bernstein}, {Bertin}, {Bridle}, {Brown}, {Clowe}, {Dahle}, {Erben}, {Gray},
  {Hetterscheidt}, {Hoekstra}, {Hudelot}, {Jarvis}, {Kuijken}, {Margoniner},
  {Massey}, {Mellier}, {Nakajima}, {Refregier}, {Rhodes}, {Schrabback}, \&
  {Wittman}}]{heymans2006}
{Heymans}, C., {Van Waerbeke}, L., {Bacon}, D., {et~al.} 2006, \mnras, 368,
  1323, \dodoi{10.1111/j.1365-2966.2006.10198.x}

\bibitem[{{Hoekstra}(2021)}]{HoekstraMdet2021b}
{Hoekstra}, H. 2021, \aap, 656, A135, \dodoi{10.1051/0004-6361/202141670}

\bibitem[{{Hoekstra} {et~al.}(2021){Hoekstra}, {Kannawadi}, \&
  {Kitching}}]{HoekstraMdet2021a}
{Hoekstra}, H., {Kannawadi}, A., \& {Kitching}, T.~D. 2021, \aap, 646, A124,
  \dodoi{10.1051/0004-6361/202038998}

\bibitem[{{Huff} \& {Mandelbaum}(2017)}]{HuffMcal2017}
{Huff}, E., \& {Mandelbaum}, R. 2017, arXiv: 1702.02600.
\newblock \doarXiv{1702.02600}

\bibitem[{Ivezi\'c {et~al.}(2008)Ivezi\'c, Tyson, Acosta, Allsman, Anderson,
  Andrew, Angel, Axelrod, Barr, Becker, {et~al.}}]{IvezicLSST2008}
Ivezi\'c, v., Tyson, J.~A., Acosta, E., {et~al.} 2008.
\newblock \doarXiv{0805.2366v4}

\bibitem[{{Kamath} {et~al.}(2020){Kamath}, {Meyers}, {Burchat}, \& {(LSST Dark
  Energy Science Collaboration}}]{Kamath2020}
{Kamath}, S., {Meyers}, J.~E., {Burchat}, P.~R., \& {(LSST Dark Energy Science
  Collaboration}. 2020, \apj, 888, 23, \dodoi{10.3847/1538-4357/ab54cb}

\bibitem[{{Laureijs} {et~al.}(2011)}]{Euclid2011}
{Laureijs}, R., {et~al.} 2011, arXiv e-prints, arXiv:1110.3193.
\newblock \doarXiv{1110.3193}

\bibitem[{{Li} {et~al.}(2022){Li}, {Kuijken}, {Hoekstra}, {Miller}, {Heymans},
  {Hildebrandt}, {van den Busch}, {Wright}, {Yoon}, {Bilicki}, {Bravo}, \&
  {Lagos}}]{LiNofz2022}
{Li}, S.-S., {Kuijken}, K., {Hoekstra}, H., {et~al.} 2022, arXiv e-prints,
  arXiv:2210.07163.
\newblock \doarXiv{2210.07163}

\bibitem[{{Li} \& {Mandelbaum}(2022)}]{LiFPFSBlending2022}
{Li}, X., \& {Mandelbaum}, R. 2022, arXiv e-prints, arXiv:2208.10522.
\newblock \doarXiv{2208.10522}

\bibitem[{Lupton(1993)}]{LuptonStats1993}
Lupton, R. 1993, Statistics in Theory and Practice (Princeton University
  Press).
\newblock \url{http://www.jstor.org/stable/j.ctvzxx986}

\bibitem[{{MacCrann} {et~al.}(2022)}]{MacCrann2022}
{MacCrann}, N., {et~al.} 2022, \mnras, 509, 3371,
  \dodoi{10.1093/mnras/stab2870}

\bibitem[{Mandelbaum {et~al.}(2018)}]{SRD}
Mandelbaum, R., {et~al.} 2018, The LSST Dark Energy Science Collaboration
  (DESC) Science Requirements Document,  arXiv,
  \dodoi{10.48550/ARXIV.1809.01669}

\bibitem[{{Massey} {et~al.}(2013){Massey}, {Hoekstra}, {Kitching}, {Rhodes},
  {Cropper}, {Amiaux}, {Harvey}, {Mellier}, {Meneghetti}, {Miller},
  {Paulin-Henriksson}, {Pires}, {Scaramella}, \& {Schrabback}}]{Massey2013}
{Massey}, R., {Hoekstra}, H., {Kitching}, T., {et~al.} 2013, \mnras, 429, 661,
  \dodoi{10.1093/mnras/sts371}

\bibitem[{Melchior {et~al.}(2018)Melchior, Moolekamp, Jerdee, Armstrong, Sun,
  Bosch, \& Lupton}]{MelchiorScarlet2018}
Melchior, P., Moolekamp, F., Jerdee, M., {et~al.} 2018, Astronomy and
  Computing, 24, 129, \dodoi{10.1016/j.ascom.2018.07.001}

\bibitem[{{Meyers} \& {Burchat}(2015)}]{MeyersBurchat2015}
{Meyers}, J.~E., \& {Burchat}, P.~R. 2015, \apj, 807, 182,
  \dodoi{10.1088/0004-637X/807/2/182}

\bibitem[{{Moffat}(1969)}]{Moffat1969}
{Moffat}, A.~F.~J. 1969, \aap, 3, 455

\bibitem[{{Plazas} \& {Bernstein}(2012)}]{PlazasPSF2012}
{Plazas}, A.~A., \& {Bernstein}, G. 2012, \pasp, 124, 1113,
  \dodoi{10.1086/668294}

\bibitem[{{Pujol} {et~al.}(2019){Pujol}, {Kilbinger}, {Sureau}, \&
  {Bobin}}]{pujol2019}
{Pujol}, A., {Kilbinger}, M., {Sureau}, F., \& {Bobin}, J. 2019, \aap, 621, A2,
  \dodoi{10.1051/0004-6361/201833740}

\bibitem[{Sanchez {et~al.}(2021)Sanchez, Mendoza, Kirkby, \&
  Burchat}]{DESCWLSanchez2021}
Sanchez, J., Mendoza, I., Kirkby, D.~P., \& Burchat, P.~R. 2021, Journal of
  Cosmology and Astroparticle Physics, 2021, 043,
  \dodoi{10.1088/1475-7516/2021/07/043}

\bibitem[{{Sheldon} {et~al.}(2020){Sheldon}, {Becker}, {MacCrann}, \&
  {Jarvis}}]{mdet20}
{Sheldon}, E.~S., {Becker}, M.~R., {MacCrann}, N., \& {Jarvis}, M. 2020, \apj,
  902, 138, \dodoi{10.3847/1538-4357/abb595}

\bibitem[{{Sheldon} \& {Huff}(2017)}]{SheldonMcal2017}
{Sheldon}, E.~S., \& {Huff}, E.~M. 2017, \apj, 841, 24,
  \dodoi{10.3847/1538-4357/aa704b}

\bibitem[{{Spergel} {et~al.}(2015)}]{Roman2015}
{Spergel}, D., {et~al.} 2015, arXiv e-prints, arXiv:1503.03757.
\newblock \doarXiv{1503.03757}

\bibitem[{{Suchyta} {et~al.}(2016)}]{SuchytaBalrog2016}
{Suchyta}, E., {et~al.} 2016, \mnras, 457, 786, \dodoi{10.1093/mnras/stv2953}

\bibitem[{{Zhang} {et~al.}(2022){Zhang}, {Becker}, \& {Sheldon}}]{dfmcal22}
{Zhang}, Z., {Becker}, M.~R., \& {Sheldon}, E.~S. 2022, arXiv e-prints,
  arXiv:2206.07683.
\newblock \doarXiv{2206.07683}

\end{thebibliography}
